\documentclass[doublecol]{epl2_mod}

\usepackage{amsmath}
\usepackage{amsfonts}
\usepackage{amssymb}
\usepackage{graphicx}
\usepackage{moreverb}
\usepackage{url}
\usepackage{epsf}
\usepackage{color}
\usepackage{multirow}
\usepackage{subfigure}

\title{Negative-U properties for substitutional Au in Si}
\shorttitle{Negative-U properties for substitutional Au in Si}

\author{Fabiano Corsetti\inst{1,2} \and Arash A. Mostofi\inst{1}}
\shortauthor{F. Corsetti \etal}

\institute{                    
  \inst{1} CIC nanoGUNE, 20018 Donostia-San Sebasti\'{a}n, Spain\\
  \inst{2} Departments of Materials and Physics, and the Thomas Young Centre for Theory and Simulation of Materials, Imperial College London, London SW7 2AZ, United Kingdom
}

\abstract{
The isolated substitutional gold impurity in bulk silicon is studied in detail using electronic structure calculations based on density-functional theory. The defect system is found to be a non-spin-polarized negative-U centre, thus providing a simple solution to the long-standing debate over the electron paramagnetic resonance signal for gold in silicon. There is an excellent agreement (within 0.03~eV) between the well-established experimental donor and acceptor levels and the predicted stable charge state transition levels, allowing for the unambiguous assignment of the two experimental levels to the $\left ( {1+} / {1-} \right )$ and $\left ( {1-} / {3-} \right )$ transitions, respectively, in contrast to previously held assumptions about the system.
}

\begin{document}

\maketitle

\section{Introduction}

The role of gold as an impurity in silicon has been studied extensively, due to its technological importance in the semiconductor industry: gold doping introduces deep defect levels into the band gap of silicon; such deep-level impurities reduce minority carrier lifetimes, and can lead to a significant decrease in the conductivity of the system~\cite{gold_minority}. It has also received renewed interest due to the use of gold as a catalyst in the growth of semiconducting nanowires using the vapor-liquid-solid (VLS) method~\cite{start-nano}, which can result in the presence of gold impurities in excess of the bulk solubility~\cite{nature-et-al}.

Since 1957, two defect levels have been identified as the main contributions from Au impurity centres: a donor level at $\varepsilon_v + 0.35 \pm 0.02$~eV, and an acceptor level at $\varepsilon_v + 0.62 \pm 0.02$~eV, where $\varepsilon_v$ is the valence band edge and the band gap is 1.16~eV~\cite{origin2level}. These measurements have been confirmed by several different experimental techniques~\cite{zunger}, including a relatively recent study~\cite{dlts-study} using deep-level transient spectroscopy (DLTS), which has clearly identified the defect centre as a single amphotheric Au substitutional. However, the electronic structure of this defect is still a matter of debate; this is mainly due to the lack of an electron paramagnetic resonance (EPR) signal~\cite{zunger}, in contrast to the isoelectronic $\mathrm{Pt}^-$ defect, which has been well characterized using EPR~\cite{pt-epr}. A possible explanation for this situation has been given by Anderson~\cite{no-epr}, who proposed that rapid tunneling of the defect between two trigonal $\mathrm{C_{2v}}$ configurations results in an average value of $g_\perp \simeq 0$. This would make the microwave transition probability between states small and hence any EPR signal difficult to detect. This explanation is supported by Watkins {\em et al.}~\cite{amp2} in Zeeman studies of the substitutional Au excitation spectra; on the other hand, Son {\em et al.}~\cite{son-au} have attributed an EPR spectrum to this defect centre, concluding that it is paramagnetic with $S = 1/2$, but with substantially different $g$ values to those found in Ref.~\cite{amp2}, which makes it unlikely that both studies are examining the same centre. In each case there is a problem with unambiguously identifying the defect centre, due to the presence of other impurities and the large number of Au-related complexes that are known to exist~\cite{au-complex,au-complex3}.

In spite of its technological relevance and the many experimental studies that have been undertaken, there have been very few computational simulations of the electronic structure of this important defect centre. Self-consistent Green's function calculations of the Au substitutional defect in an undistorted Si lattice~\cite{zunger} have shown the possibility of both the donor and acceptor levels arising from the same centre, in contrast to the results from a previous cluster model~\cite{alpha-et-al}; the predicted level positions, however, are 0.21~eV and 0.26~eV higher than the experimental measurements for the donor and acceptor levels, respectively. More recently, a density-functional theory (DFT) cluster simulation~\cite{new-half} using a relaxed ionic configuration also obtained two levels, although their positions were found to be 0.14~eV and 0.12~eV lower than the experimental donor and acceptor levels, respectively, in spite of the application of an empirical correction.

In this Letter, we determine accurately the donor and acceptor levels of the isolated Au substitutional defect in bulk Si using large-scale first-principles DFT calculations. It is notable that our theoretical predictions, calculated without the use of any empirical corrections or reference levels, are found to be in excellent agreement with experiment (within 0.03~eV): $\varepsilon_v+0.37$~eV for the donor level, and $\varepsilon_v+0.59$~eV for the acceptor level. These levels correspond to the stable charge state transitions $\left ( {1+} / {1-} \right )$ and $\left ( {1-} / {3-} \right )$, showing the neutral and doubly negative charge states to be unstable. The defect centre, therefore, exhibits a negative-U effect~\cite{negU-baraff,negU} and is not paramagnetic, thereby explaining the absence of an EPR signal.

Our calculations represent a significant improvement on previous theoretical studies of this defect centre, for the following reasons: (i) a broader range of charge states has been investigated (from $1+$ to $3-$), (ii) a large supercell is used, and (iii) careful consideration has been given to the sampling of electronic states in the Brillouin zone (BZ), so that the sampled states possess the full point group symmetry of an isolated defect in an infinite crystal. We note that previous computational studies of the finite size convergence properties of the silicon vacancy~\cite{niem-vac,wright-vac,us-vac} (a Jahn-Teller defect system closely related to the gold centre~\cite{vac}) indicate that points (ii) and (iii) are crucial for obtaining both the correct relaxation of the ions around the defect centre, and the accurate positioning of the transition levels in the band gap.

\section{Computational methods}

The formation energy of the Au substitutional defect centre in charge state $q$ (denoted $\mathrm{Au}^q$) is defined as
\begin{equation} \label{eqn:E_f}
E_f^q = E_\mathrm{def}^q - \left ( \frac{N-1}{N} \right ) E_\mathrm{bulk} - \mu_\mathrm{Au} + q \mu_e,
\end{equation}
where $E_\mathrm{bulk}$ is the total energy of the bulk supercell with $N$ atoms, $E_\mathrm{def}^q$ is the total energy of the same supercell with one substitutional defect, $\mu_\mathrm{Au}$ is the chemical potential of the gold atom species\footnote{$\mu_\mathrm{Au}$ is taken to be the energy per atom of ideal FCC gold calculated with the LSDA-DFT-optimized lattice parameter of 4.04~$\mathrm{\AA}$, which is within 1\% of the experimental value of 4.08~$\mathrm{\AA}$.}, and $\mu_e$ is the electronic chemical potential. This last term is calculated from the energy difference of the neutral bulk supercell and the same supercell with a single electron hole~\cite{us-vac}. In the remainder of this Letter we shall refer to $\Delta \mu_e$ as the electronic chemical potential relative to $\varepsilon_v$.

The calculations are performed using the plane-wave pseudopotential DFT code {\sc castep}~\cite{castep-et-al} (version 5.0). Unless otherwise stated, we use a 256-atom BCC supercell of the host material (bulk silicon), with $\Gamma$-point sampling of the BZ. Exchange and correlation (xc) is treated within the Ceperley-Alder local-spin-density approximation~\cite{qmc1} (LSDA). Core electrons are pseudized using the Vanderbilt ultrasoft pseudopotential formalism~\cite{ultra-pseudo}, with 4 and 19 valence electrons for Si and Au, respectively. A plane-wave cut-off energy of 400~eV is used, which is sufficient to converge defect formation energies to within a tolerance of 10~meV. The relaxed bulk Si lattice constant, used for all of our calculations, is 5.39~$\mathrm{\AA}$, which is within 1\% of the experimental value of 5.43~$\mathrm{\AA}$. When calculating relaxed geometries, all the ions in the supercell are allowed to move, and their initial positions are given small randomized displacements to allow for symmetry breaking; our convergence tolerance is 6~meV/$\mathrm{\AA}$ for the maximum force on each ion.

\begin{figure*}
\begin{center}
\subfigure[Unrelaxed lattice]{\includegraphics{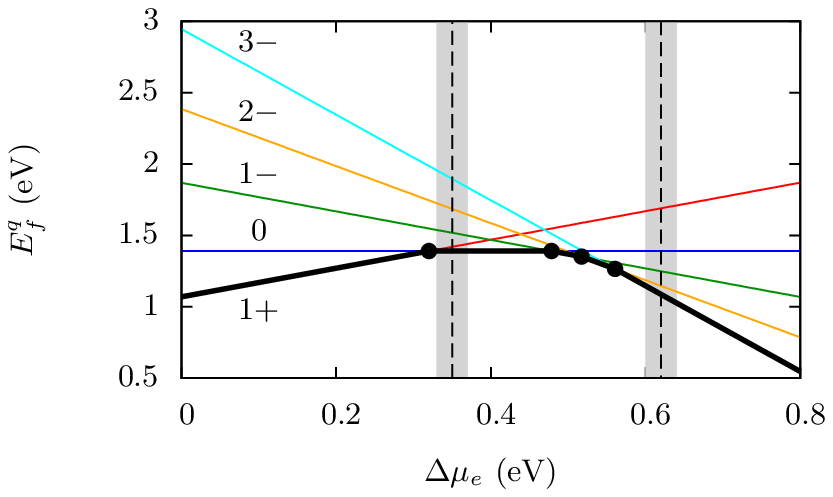}}
\subfigure[Relaxed lattice]{\includegraphics{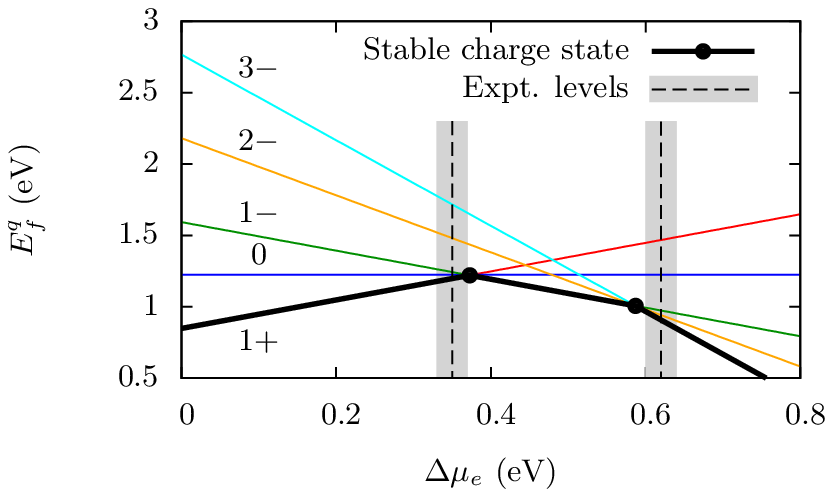}}
\caption{Formation energy of the different charge states of the defect as a function of the electronic chemical potential (plotted relative to the VBM) using LSDA-DFT. The complete band gap (not shown) is of 1.2~eV. The thermodynamically stable charge state at each point is highlighted in bold, and the circles indicate the level position for the stable transitions. The dashed vertical lines show the experimental values for the transition levels (the uncertainty range is shown in gray).\label{fig:stable}}
\end{center}
\end{figure*}

\section{Results}

\begin{table}
\caption{Transition levels (in eV) obtained for relaxed (Rel.) and unrelaxed (Unrel.) geometries using LSDA-DFT. Results from previous studies are also shown. Asterisks (*) denote thermodynamically stable transitions. Experimental values for the donor and acceptor levels are placed in line with our proposed corresponding transition levels, as explained in the text.}
\label{table:levels}
\begin{tabular}{lccccc}
\hline
Ref.  & \cite{origin2level} & \multicolumn{2}{c}{This study} & \cite{zunger} & \cite{new-half} \\ \cline{3-4}
      & Expt.               & Rel. & Unrel.                  & Unrel.        & Rel.            \\
\hline
$E \left ( {1+} / 0 \right )$    & --   & {\color{white}*}0.38{\color{white}*} & {\color{white}*}0.32*                & 0.56 & 0.21 \\
$E \left ( {1+} / {1-} \right )$ & 0.35 & {\color{white}*}0.37*                & {\color{white}*}0.40{\color{white}*} & --   & --   \\
$E \left ( 0 / {1-} \right )$    & --   & {\color{white}*}0.37{\color{white}*} & {\color{white}*}0.48*                & 0.88 & 0.50 \\
$E \left ( 0 / {2-} \right )$    & --   & {\color{white}*}0.48{\color{white}*} & {\color{white}*}0.50{\color{white}*} & --   & --   \\
$E \left ( {1-} / {2-} \right )$ & --   & {\color{white}*}0.59{\color{white}*} & {\color{white}*}0.52*                & --   & --   \\
$E \left ( {1-} / {3-} \right )$ & 0.62 & {\color{white}*}0.59*                & {\color{white}*}0.54{\color{white}*} & --   & --   \\
$E \left ( {2-} / {3-} \right )$ & --   & {\color{white}*}0.59{\color{white}*} & {\color{white}*}0.56*                & --   & --   \\
\hline
\end{tabular}
\end{table}

Following the definition of Baraff {\em et al.}~\cite{negU-baraff}, the stable charge state transition level $E \left ( m / n \right )$ is the value of $\Delta \mu_e$ at which there is a crossing of the defect formation energies of two charge states $m$ and $n$, leading to a change in the most stable state from $\mathrm{Au}^m$ to $\mathrm{Au}^n$ as the Fermi level is raised. Our transition levels are given in Table~\ref{table:levels} for both the relaxed and unrelaxed lattice.

The ionic relaxation has a large effect on the level ordering and positions, as shown in Fig.~\ref{fig:stable}. For the unrelaxed lattice, all the charge states from $1+$ to $3-$ exhibit a thermodynamically stable region, with the spin-polarized neutral charge state having the lowest defect formation energy for a range of $\Delta \mu_e$ of width 0.16~eV. The lattice relaxation, however, significantly lowers the formation energy of $\mathrm{Au}^{1-}$ (by between 0.07~eV and 0.16~eV more than the other charge states), resulting in both the neutral and doubly negative charge states being cut off, and two thermodynamically stable double electron transitions appearing, from $\mathrm{Au}^{1+}$ to $\mathrm{Au}^{1-}$, and from $\mathrm{Au}^{1-}$ to $\mathrm{Au}^{3-}$. In other words, the defect centre is a negative-effective-U system as a direct consequence of the Jahn-Teller lattice distortion, analogously to the behavior observed for the Si vacancy~\cite{negU}. 

The position of the two thermodynamically stable transition levels for the relaxed system $E \left ( {1+} / {1-} \right )$ and $E \left ( {1-} / {3-} \right )$ are in good agreement with the donor and acceptor levels measured experimentally, with a discrepancy of $+0.02$~eV and $-0.03$~eV, respectively. We note that experimental measurements also vary on the order of 0.01~eV~\cite{origin2level,zunger,dlts-study}.

\begin{table}
\caption{Geometry of the defect centre after relaxation for the 256-atom supercell using LSDA-DFT. Bond lengths are given between pairs of Si ions $\mathrm{a}, \mathrm{b}$ surrounding the defect centre (Fig.~\ref{fig:Au_Si_watkins}) and the Au impurity ion $\mathrm{c}$. Equivalent bonds in each system agree to within 0.01~$\mathrm{\AA}$, except for ones labeled with a dagger ($\dagger$), which agree to within 0.03~$\mathrm{\AA}$. The defect volume is calculated from the tetrahedron formed by the four neighbours of the impurity. The last line is for the unrelaxed centre.}
\label{table:summary}
\begin{tabular}{ccccccccc}
\hline
                     &                                       & Vol.                  & \multicolumn{5}{c}{Bond lengths ($\mathrm{\AA}$)} \\ \cline{4-8}
$q$                  & Sym.                                  & ($\mathrm{\AA}^3$)    & a--a                  & b--b                  & a--b                          & a--c                  & b--c                  \\
\hline
$1+$                 & $\mathrm{D_{2d}}$                     & 6.78                  & 3.71                  & 3.71                  & $3.95{\color{white}^\dagger}$ & 2.37                  & 2.37                  \\
\multirow{2}{*}{$0$} & \multirow{2}{*}{$\sim$$\mathrm{D_2}$} & \multirow{2}{*}{6.72} & \multirow{2}{*}{4.00} & \multirow{2}{*}{3.98} & $3.86/$                       & \multirow{2}{*}{2.36} & \multirow{2}{*}{2.37} \\
                     &                                       &                       &                       &                       & $3.73{\color{white}^\dagger}$ &                       &                       \\
$1-$                 & $\sim$$\mathrm{D_{2d}}$               & 6.60                  & 4.02                  & 4.02                  & $3.75^\dagger$                & 2.35                  & 2.35                  \\
$2-$                 & $\sim$$\mathrm{D_{2d}}$               & 6.60                  & 4.02                  & 4.01                  & $3.75^\dagger$                & 2.35                  & 2.35                  \\
$3-$                 & $\sim$$\mathrm{D_{2d}}$               & 6.61                  & 4.03                  & 4.00                  & $3.76^\dagger$                & 2.35                  & 2.36                  \\
\hline
                     & $\mathrm{T_d}$                        & 6.54                  & 3.81                  & 3.81                  & $3.81{\color{white}^\dagger}$ & 2.34                  & 2.34                  \\
\hline
\end{tabular}
\end{table}

\begin{figure}
\begin{center}
\includegraphics[width=0.49\textwidth]{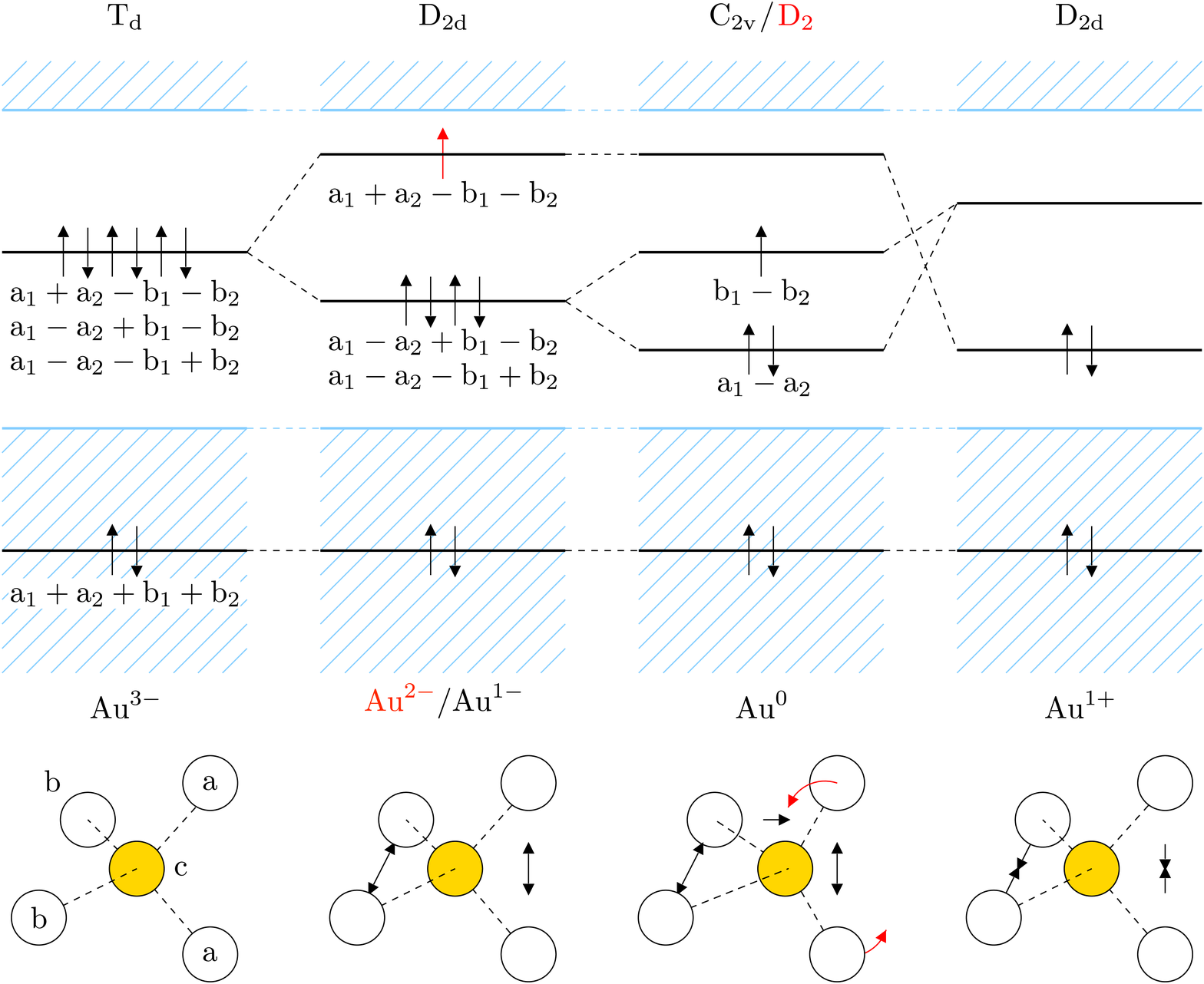}
\caption{Effect of Jahn-Teller distortion on the defect centre for different charge states, following Watkins' LCAO model. The four orbitals a$_1$, a$_2$, b$_1$, b$_2$ are hybrids of the dangling bonds surrounding the defect and the 6sp$^3$ valence orbitals of Au~\cite{zunger}.\label{fig:Au_Si_watkins}}
\end{center}
\end{figure}

The ionic structure of the relaxed defect centre is given in Table~\ref{table:summary}. The observed relaxation patterns can be understood using Watkins' linear combination of atomic orbitals (LCAO) model of the vacancy~\cite{watkins-model} and transition element impurities~\cite{vac,watkins-metals} in Si, as shown in Fig.~\ref{fig:Au_Si_watkins}. In our case, $\mathrm{D_{2d}}$ symmetry is obtained via either the removal or addition of an electron from $\mathrm{Au}^0$; this is achieved by forming pairs between the four neighbouring atoms of the defect centre, although in the former case the distance between pairs decreases, whilst in the latter it increases. For $\mathrm{Au}^0$ the degeneracy is lifted on all the defect states; this is conventionally assumed to be caused by a second distortion which differentiates between the pairs, resulting in $\mathrm{C_{2v}}$ symmetry. Our calculations show some evidence of such behavior, as the two stretched bonds between ion pairs differ in length for this charge state, and the Au ion moves along [100] towards the centre of mass of the longer of the two (similarly to the findings of a previous study~\cite{new-half}). However, we also find a much more prominent distortion: a rotation of the ion pairs about the [100] axis which breaks the symmetry of the four bonds between pairs of neighbours (the a--b column in Table~\ref{table:summary}). This distortion is sufficient in itself to reduce the symmetry of the defect centre to $\mathrm{D_2}$ and lift the degeneracy of the three defect states.

Almost no difference is observed in the relaxed ionic structure of the defect centre for charge states from $1-$ to $3-$, in contrast to the model's prediction of $\mathrm{T_d}$ symmetry for $3-$. However, this result is in agreement with what has been reported for the Si vacancy~\cite{niem-vac,wright-vac,us-vac}, for which the change in the relaxed defect volume monotonically decreases as more electrons are added, and a deviation from the model's symmetry predictions is observed for significant negative charge. The stability of the double electron transition $E \left ( {1-} / {3-} \right )$ upon relaxation is, therefore, due to the relative differences in energy of the three charge states for a common change in symmetry, rather than to distinct relaxed symmetries, as is more commonly the case.

We note that both for the $\left ( {1+} / {1-} \right )$ and $\left ( {1-} / {3-} \right )$ transitions the change in the total charge density upon addition of the two extra electrons is localized predominantely in the first few shells around the defect centre. We give a detailed analysis of the electronic structure of the system using maximally-localized Wannier functions in Sec.~6.4.5 of Ref.~\cite{fcthesis}, confirming the validity of the LCAO model.

\section{Discussion}

The results presented in the previous section are obtained with a $\Gamma$-point BZ sampling. As already noted, our aim is to correctly describe the symmetry-breaking Jahn-Teller distortion of the truly isolated defect centre, whilst working within the constraints of a periodically-repeated supercell. After relaxation, any substitutional impurity in a diamond lattice can have a point group symmetry up to and including $\mathrm{T_d}$, and, hence, any sampled k-point should possess all elements of this group. Amongst the high-symmetry points of the three cubic Bravais lattices, there are only three k-points other than $\Gamma$ that possess the required degree of symmetry: $\mathrm{R} = \left ( \frac{1}{2}, \frac{1}{2}, \frac{1}{2} \right )$ for the simple cubic (SC) lattice, and $\mathrm{H} = \left ( \frac{1}{2}, \frac{1}{2}, -\frac{1}{2} \right )$ and $\mathrm{P} = \left ( \frac{1}{4}, \frac{1}{4}, \frac{1}{4} \right )$ for the BCC lattice\footnote{The high-symmetry points are given as fractional coordinates of the reciprocal lattice vectors for each Bravais lattice.}; there are none for the FCC lattice. R and H have $\mathrm{O_h}$ symmetry, and P has $\mathrm{T_d}$ symmetry. Previous studies have indeed suggested the use of different combinations of these k-points for substitutional defects in cubic host lattices~\cite{niem-vac,sampling_si-et-al,positron_defects}. 

\begin{figure}
\begin{center}
\subfigure[DFT calculations]{\includegraphics{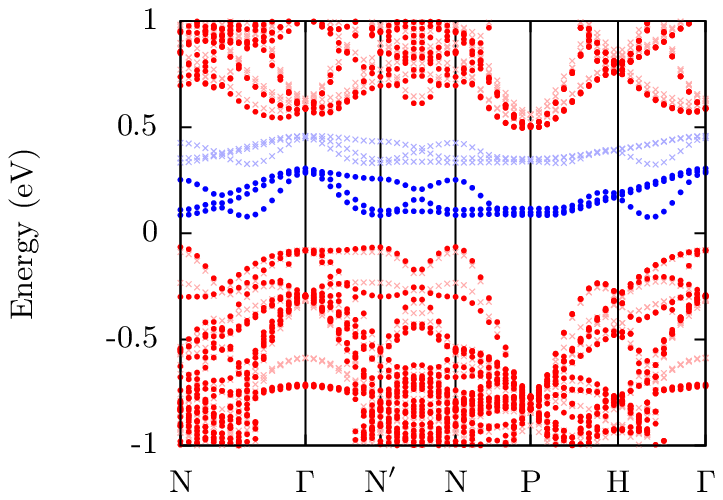}}
\subfigure[Tight-binding model]{\includegraphics{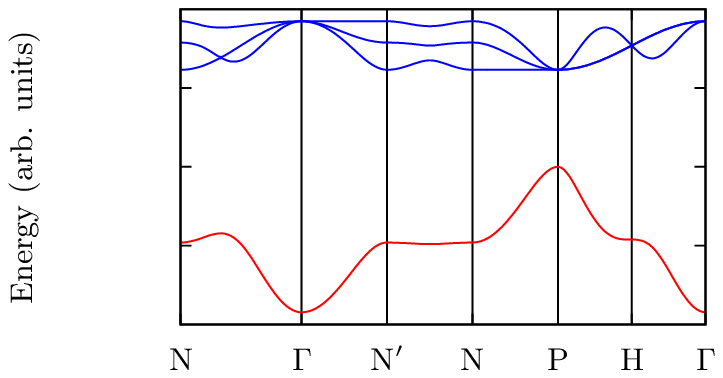}}
\caption{Band structure of the undistorted BCC defect supercell from (a) LSDA-DFT and (b) the tight-binding model described in the Appendix. The three defect bands in the band gap are shown in blue. For (a), the calculation is of the 256-atom supercell, with dark filled circles showing the Au substitutional system and light crosses the vacancy (both in their neutral charge state). For (b), the red band is the nodeless combination buried within the bulk Si valence band, as shown in Fig.~\ref{fig:Au_Si_watkins}.}
\label{fig:extended_bs}
\end{center}
\end{figure}

In order to assess the effect of k-point sampling on the properties of the defect supercell, we have developed a tight-binding model of the system that extends Watkins' four-orbital LCAO model to include a small hopping matrix element between defect centres in neighbouring supercells (see the Appendix for a description of the model). In Fig.~\ref{fig:extended_bs} it can be seen that this model accurately describes the qualitative details of the three disperse defect bands in the band gap as compared with DFT calculations of both the gold centre and the vacancy. Furthermore, the model predicts an exact cancellation of all finite size effects (terms depending on the inter-supercell hopping element) at the high-symmetry points $\mathrm{R}$, $\mathrm{H}$, $\mathrm{P}$ and $\Gamma$, so that the degree of Jahn-Teller distortion and corresponding decrease in energy is the same as for an isolated defect centre. Instead, the use of a dense grid that samples low-symmetry k-points favors the undistorted lattice, resulting in no distortion below a critical system size; this behavior has previously been found in DFT calculations of the vacancy~\cite{niem-vac,us-vac}. The model shows that above this critical value the correct distortion is recovered even for low-symmetry k-point samplings, although the corresponding decrease in energy is underestimated with respect to the isolated defect limit. We believe that these observations explain the poor finite size convergence of the transition levels reported for the vacancy when using a dense k-point grid instead of $\Gamma$-point only sampling~\cite{us-vac}; we postulate that this is a general phenomenon for Jahn-Teller defects in diamond lattices.

Due to the simplicity of the tight-binding model, however, we do not expect it to give quantitatively accurate predictions for the relative finite size convergence of the various high-symmetry k-points. Therefore, we have calculated with DFT the transitions for the sequence $\mathrm{Au}^{1-}$, $\mathrm{Au}^0$, $\mathrm{Au}^{1+}$ with $\Gamma + \mathrm{H}$ sampling~\cite{positron_defects} and $\mathrm{P}$-point only sampling; in both cases, the negative-U effect is maintained, such that $E \left ( {1+} / {1-} \right )$ is the only thermodynamically stable transition. The position of this level, however, changes significantly with respect to the $\Gamma$-point only sampling case (by $-0.09$~eV and $+0.18$~eV, respectively). In order to check which of these sampling schemes gives the most accurate level positions, we have performed the calculation using a large 864-atom BCC supercell, by embedding the relaxed ionic positions around the defect site obtained from the 256-atom calculation into the larger system\footnote{The root mean square residual ionic force in these supercells is less than 0.1~eV/$\mathrm{\AA}$.}. The transition level obtained is at $\varepsilon_v + 0.34$~eV; among the different sampling schemes tested in the 256-atom system, $\Gamma$-point only has the smallest discrepancy with respect to this value ($+0.03$~eV).

Another important question relates to the accuracy of the LSDA for exchange and correlation, since it is well known that both local and semi-local functionals suffer from a pathological self-interaction error~\cite{delta^convex}, which can be problematic for correctly placing deep defect levels in the fundamental band gap~\cite{def_corr}. Many correction methods have been proposed~\cite{def_corr}, notably the use of non-local hybrid functionals~\cite{hybrid2,hybrid1,Spiewak2013}. We have therefore performed additional tests on our system using the HSE06 hybrid functional~\cite{xc-HSE06}. Since we are mainly interested in checking discrepancies with the LSDA, we use the smaller 32-atom BCC supercell with $\Gamma$-point only sampling, which exhibits the same qualitative features of the negative-U defect centre already described. For the HSE06 calculations, we employ norm-conserving pseudopotentials and a plane-wave cut-off energy of 500~eV. The relaxed geometries are very close to those predicted by the LSDA (Table~\ref{table:summary}): the two functionals agree on the symmetry of the defect centre for all charge states with the exception of $\mathrm{Au}^{3-}$, which reverts to $\sim$$\mathrm{D_2}$ with HSE06. Equivalent relaxations with the LSDA on this smaller 32-atom supercell, however, also give a different low symmetry configuration for this charge state only, suggesting a finite-size effect.

Most importantly, HSE06 also predicts a negative-U effect for both transitions. There is, however, a noticeable difference in the placement of the two transition levels in the band gap, with HSE06 predicting $E \left ( {1+} / {1-} \right )$ and $E \left ( {1-} / {3-} \right )$ to be 0.43 eV and 0.50 eV higher, respectively, than with the LSDA. Such a shift is in qualitative agreement with previous predictions~\cite{hybrid2}; however, in view of the somewhat empirical nature of the correction offered by the hybrid functional, we should not necessarily {\em a priori} expect an improvement over the LSDA in the level placements. Indeed, previous studies have already provided a compelling case for the accuracy of the LSDA in describing Jahn-Teller substitutional defects in Si: the transition levels for the vacancy are predicted within a few meV of experiment~\cite{negU,us-vac}. Using identical methods we find the same level of agreement for the Au defect. A possible explanation for this is that, despite being `deep' from a quantitative perspective, the behavior of the defect levels is actually more similar to that of a shallow level, tied to the valence band edge (type-I behavior, as described in Ref.~\cite{def_corr}); this is also confirmed by the simplified picture offered by Watkins' model, in which four defect states originate from perturbed valence orbitals. In Fig.~\ref{fig:stable}, therefore, we show the transition levels within the experimental band gap; this effectively equates to applying a scissor operator correction scheme~\cite{Levine1989} so as to shift the conduction bands to higher energies.

Finally, we have tested for additional effects due to spin-orbit coupling\footnote{Calculations were performed with the {\sc abinit} code~\cite{abinit-generic-et-al} (version 6.6), the LDA for xc, HGH norm-conserving pseudopotentials~\cite{hgh-pseudo}, $\Gamma$-point only sampling, and a plane-wave cut-off energy of 800~eV.}, the inclusion of which gives rise to a shift of approximately $-0.1$~eV in the defect formation energy. When considering transition levels, this shift almost entirely cancels between different charge states, resulting in negligible changes to the transition levels ($E \left ( {1+} / {1-} \right )$ and $E \left ( {1-} / {3-} \right )$ being lowered by only 15~meV and 17~meV, respectively) and the relaxed ionic configurations, and no qualitative change in the negative-U nature of the defect.

\section{Conclusions}

We have proposed a solution to the important question of the missing EPR signal of gold in silicon: this point defect system is a negative-U centre (similarly to the vacancy) due to a significant Jahn-Teller effect, that favors the spin-unpolarized odd charge states (most of all $\mathrm{Au}^{1-}$) over the spin-polarized even ones, thereby causing the latter to be thermodynamically unstable. The robustness of our results is supported by the quantitative agreement, within chemical accuracy, between the two experimental defect levels and our simulated predictions. Spin-orbit coupling is found to be a comparatively weak effect that has no qualitative bearing on the main conclusion of this study. Additional tests using a hybrid functional for describing exchange and correlation also confirm the double negative-U effect, although the two transition levels are shifted higher in the band gap.

\acknowledgments
This work was supported by the UK's EPSRC, HPC Materials Chemistry Consortium (EP/F067496/1), and Car-Parrinello Consortium (EP/K013564/1). The calculations were performed on Imperial College HPC and the UK's HECToR. We thank W.~M.~C.~Foulkes and P.~D.~Haynes for helpful discussions.

\section{Appendix: The extended LCAO model}

\begin{widetext}
\vspace{-10pt}
\begin{equation} \label{eqn:extend_H}
H^{\left ( \Delta,c \right ) } = \begin{pmatrix}
-a & \mathrm{c.c.} & \mathrm{c.c.} & \mathrm{c.c.} \\
-b-c (\mathrm{e}^{\mathrm{i} \mathbf{k} \cdot \mathbf{a}_1} + \mathrm{e}^{\mathrm{i} \mathbf{k} \cdot \mathbf{a}_4} )-\Delta & -a & \mathrm{c.c.} & \mathrm{c.c.} \\
-b-c (\mathrm{e}^{\mathrm{i} \mathbf{k} \cdot \mathbf{a}_2} + \mathrm{e}^{\mathrm{i} \mathbf{k} \cdot \mathbf{a}_4} ) & -b-c (\mathrm{e}^{-\mathrm{i} \mathbf{k} \cdot \mathbf{a}_1} + \mathrm{e}^{\mathrm{i} \mathbf{k} \cdot \mathbf{a}_2} ) & -a & \mathrm{c.c.} \\
-b-c (\mathrm{e}^{\mathrm{i} \mathbf{k} \cdot \mathbf{a}_3} + \mathrm{e}^{\mathrm{i} \mathbf{k} \cdot \mathbf{a}_4} ) & -b-c (\mathrm{e}^{-\mathrm{i} \mathbf{k} \cdot \mathbf{a}_1} + \mathrm{e}^{\mathrm{i} \mathbf{k} \cdot \mathbf{a}_3} ) & -b-c (\mathrm{e}^{-\mathrm{i} \mathbf{k} \cdot \mathbf{a}_2} + \mathrm{e}^{\mathrm{i} \mathbf{k} \cdot \mathbf{a}_3} )-\Delta & -a
\end{pmatrix}
\end{equation}\end{widetext}

We briefly describe here our extension to Watkins' LCAO model of the defect~\cite{watkins-model,vac,watkins-metals}. A more detailed discussion of our model is presented in Appendix~D of Ref.~\cite{fcthesis}.

Assuming that the defect does not interact with the host lattice, we use the four dangling bonds of the Si neighbours of the defect site to make the tight-binding Hamiltonian given in Eq.~\ref{eqn:extend_H}. The parameters $a$ and $b$ are the absolute values of the on-site and two-centre matrix elements for the undistorted lattice. We add a term $\Delta=\alpha Q$ to four off-diagonal elements, proportional to a hypothetical tetragonal distortion $Q$ of the system; $Q$ is related to the bond length between a pair of dimerized neighbours of the defect centre, where $Q=0$ corresponds to the undistorted bond length, and $Q=1$ to the limit of zero bond length. Finally, $c$ is the strength of the interaction with the nearest orbitals from defect images in neighbouring supercells (in the simplest approximation for BCC, each orbital interacts with six extra orbitals). The translations are written in terms of the supercell lattice vectors $\left \{ \mathbf{a}_1, \mathbf{a}_2, \mathbf{a}_3\right \}$ and their sum $\mathbf{a}_4$.

Using Eq.~\ref{eqn:extend_H}, the equilibrium distortion $Q^\mathrm{eq}$ and the corresponding {\em decrease} in formation energy $\Delta E_f$ from the undistorted ($Q=0$) to the distorted ($Q=Q^\mathrm{eq}$) lattice can be calculated for different k-point sampling schemes. We are particularly interested in the influence of finite size effects ($c \neq 0$) with respect to the limit of a perfectly isolated defect in an infinite lattice ($c=0$).

We find that finite size effects cancel at high-symmetry points ($\Gamma$, $H$ and $P$ for BCC), and so both $Q^\mathrm{eq}$ and $\Delta E_f$ are identical to the isolated defect centre for all supercell sizes. The model, therefore, provides a simple explanation as to why $\Gamma$-point only calculations exhibit the correct relaxation even for small supercells~\cite{niem-vac,us-vac}.

However, this cancellation does not occur for low-symmetry k-points. As an example, we numerically investigate the behavior of the system using a $3 \times 3 \times 3$ k-point grid. There are only two independent parameters, the ratios $c/b$ and $\alpha/b$. We find that the model in this case features a discontinuity in $Q^\mathrm{eq}$: below a critical value of $c/b$ (which depends on $\alpha/b$), the correct relaxation is obtained, but above it the system remains completely undistorted. This is consistent with our observation of an abrupt change in symmetry for the neutral vacancy~\cite{us-vac}, which remains $\mathrm{T_d}$ for supercells $<$256 atoms and switches to $\mathrm{D_{2d}}$ for those $\ge$256 atoms. Furthermore, $\Delta E_f$ is zero in the undistorted region, and increases approximately linearly in the distorted region to the correct value at $c=0$. This suggests that small supercells with dense k-point sampling will underestimate the gain in energy from Jahn-Teller distortion, even if they show the correct change in symmetry.

\bibliographystyle{eplbib}

\end{document}